\documentclass[twocolumn,prb,superscriptaddress,a4paper,floatfix,showkeys]{revtex4} 
\usepackage{amsfonts}
\usepackage{amsmath}
\usepackage{amssymb}
\usepackage{graphicx}

\providecommand{\U}[1]{\rule{.1in}{.1in}}

\begin{document}

\preprint{}
\title{Event-by-event Simulation of Quantum Cryptography Protocols\footnote{To appear in J. Comput. Theor. Nanosci.}}
\author{Shuang Zhao}
\email{s.zhao@rug.nl}
\affiliation{Department of Applied Physics, Zernike Institute for Advanced Materials,
University of Groningen, Nijenborgh 4, NL-9747 AG Groningen, The Netherlands}
\author{Hans De Raedt}
\email{h.a.de.raedt@rug.nl}
\affiliation{Department of Applied Physics, Zernike Institute for Advanced Materials,
University of Groningen, Nijenborgh 4, NL-9747 AG Groningen, The Netherlands}

\begin{abstract}
We present a new approach to simulate
quantum cryptography protocols using event-based processes.
The method is validated by simulating the BB84 protocol and the Ekert
protocol, both without and with the presence of an eavesdropper.
\end{abstract}

\accepted[Accepted ]{13 August 2007}
\date{\today }
\maketitle

\def\sumprime{\mathop{{\sum}'}}

\section{Introduction}

Cryptography is an artifice of exchanging information between two parties
such that an unauthorized person cannot retrieve this information.
To this end, the sender usually employs some key to encrypt the information
to be transmitted, and the receiver applies a decryption algorithm  to recover the
original information.
If the cryptographic system is secure, an eavesdropper can decipher
the encrypted message if and only if the eavesdropper knows the key.
Thus, the central problem of cryptography is to establish a powerful key.
We may imagine that the more bits the key contains and
the more complicated the key is, the more secure the process is.
But, in practice, if the key is generated and transmitted in a
conventional, electronic way, it may be possible to intercept the key.
Then, the eavesdropper can make a copy of the exchanged information without changing
it, such that the sender or receiver did not notice that the information has been intercepted.

Quantum cryptography uses microscopic objects such as individual photons as information carriers~\cite{QCintroduction}.
One of the characteristic features of such microscopic systems is
that a measurement may change the information that the microscopic system carries.
Therefore, if an eavesdropper attempts to make a measurement to
determine a bit of the key, there is no guarantee that
the information carried by the microscopic system is left unchanged.
If the quantum cryptography protocol is designed properly,
the presence of the eavesdropper is revealed
by an increase of the error rate in the bits that
are being transmitted from sender to receiver.

Although there is no doubt that quantum theory is very successful
in describing a vast number of experimental results,
it is well-known that quantum theory has nothing to say about
individual events that are being recorded in experiments~\cite{HOME97,BALL03}.
Yet, quantum cryptography uses individual events to transmit information, its
security being guaranteed by axioms.
Since the inception of quantum theory, the major fundamental problem of incorporating in quantum theory
the fact that we observe events only is often referred to as the quantum
measurement paradox and has not yet found a solution within the realm of quantum theory~\cite{HOME97}.
Therefore, it seems worthwhile the study the fundamental question
what it is that makes quantum cryptography work: Logically speaking, it cannot be quantum mechanics because
quantum mechanics has nothing to say about individual events~\cite{HOME97,BALL03}.

In a number of recent papers~\cite{RAED05b,RAED05c,RAED05d,MICH05,RAED06c,RAED07b,ZHAO07a,RAED07c},
we have demonstrated that locally-connected networks of processing units can simulate, event-by-event,
the single-photon beam splitter, Mach-Zehnder interferometer
experiments of Grangier \textit{et al}.~\cite{GRAN86},
and Einstein-Podolsky-Rosen experiments with photons~\cite{WEIH98,WEIH00}.
Furthermore, we have shown that this approach can be generalized to simulate
universal quantum computation~\cite{NIEL00} by an event-by-event process~\cite{RAED05c,MICH05}.
Therefore, this suggests that at least in principle, it may be possible to simulate
all wave interference phenomena and many-body quantum systems using particle-like processes only.
In this paper, we extend this approach to quantum cryptography systems.

The paper is structured as follows.
In Sections~\ref{BB84} and \ref{EKERT}, we briefly review the two most popular
protocols of quantum cryptography: the BB84 protocol~\cite{BB84ref1,BB84ref2,BB84ref3}
and Ekert's protocol~\cite{Ekertref}.
Although both protocols are closely related~\cite{QCintroduction},
from the point of view of simulation algorithm, the latter is considerably more complicated
than the former, which is the main reason for discussing both of them.
The simulation algorithms for the polarizer are presented in Section~\ref{POL}.
Simulation results for the BB84 and the Ekert protocol,
both with and without eavesdropper, are given in Sections~\ref{BB84SIM} and \ref{EKERTSIM}.
Although our simulation method does not rely on any concept of quantum theory, it is nevertheless capable
of simulating quantum cryptography protocols on the event-by-event level.
Therefore, it may shine light on the question what is essential for quantum cryptography to work.
In Section~\ref{BB84SIMD}, we discuss this issue.
The results of this paper are summarized in Section~\ref{SUM}.

\begin{table*}[tp]
\caption{Summary of the BB84 protocol.
The first column shows the bit that Alice wants to encode and send to Bob.
The second and third columns give the orientations of Alice's and Bob's polarizers, respectively.
The fourth and fifth columns are the probabilities that Bob detects a photon in the output channel $0$ and $1$
of his polarizer.
The last column gives the bit that Bob obtains from his measurement.
The question mark indicates that the probability that Bob makes the wrong guess is $50\%$.}
\label{table1}
\begin{center}
\begin{ruledtabular}
\begin{tabular}
[c]{cccccc}
Alice's bit & $\phi_{A}$ & $\phi_{B}$ & $
\begin{array}
[c]{c}%
P_{0}\\
\cos^{2}(\phi_{A}-\phi_{B})
\end{array}
$ & $%
\begin{array}
[c]{c}%
P_{1}\\
\sin^{2}(\phi_{A}-\phi_{B})
\end{array}
$ & Bob's bit\\ \hline
$0$ & $0^{0}$ & $0^{0}$ & $1$ & $0$ & $0$\\
&  & $45^{0}$ & $1/2$ & $1/2$ & $?$\\
$0$ & $45^{0}$ & $0^{0}$ & $1/2$ & $1/2$ & $?$\\
&  & $45^{0}$ & $1$ & $0$ & $0$\\\hline
$1$ & $90^{0}$ & $0^{0}$ & $0$ & $1$ & $1$\\
&  & $45^{0}$ & $1/2$ & $1/2$ & $?$\\
$1$ & $135^{0}$ & $0^{0}$ & $1/2$ & $1/2$ & $?$\\
&  & $45^{0}$ & $0$ & $1$ & $1$\\
\end{tabular}
\end{ruledtabular}
\end{center}
\end{table*}

\section{BB84 protocol}
\label{BB84}

In 1989, an experimental prototype that implemented
the quantum cryptography protocol BB84~\cite{BB84ref1,BB84ref2,BB84ref3},
demonstrated that it is possible to transmit an encryption key
using the polarization state of single photons.
In this section, we briefly review the idea behind this protocol.

The BB84 protocol employs the polarization state of photons
as the information carrier.
As each detection of a single photon yields one out of two definite
answers for the polarization of the photon,
these observations can be described by the quantum theory of
a two-state system.
The BB84 protocol uses two sets of non-orthogonal coordinate systems,
which are the usual $x$-$y$ (rectilinear) basis and
the diagonal basis which is the rectilinear basis rotated by $45^\circ$.
In
the rectilinear basis, the photon can be either in the horizontal $%
(\longrightarrow )$ or in the vertical $(\uparrow )$ polarization state. In
the diagonal linear basis, the photon can be either in the diagonal $%
(\nearrow )$ or in the anti-diagonal $(\nwarrow )$ polarization state.

Let Alice and Bob be the two parties who want to
exchange a secret key. Alice generates and sends Bob a sequence of
photons with polarization states that are selected randomly from the four
possible directions: $0^{0}(\longrightarrow),45^{0}(\nearrow),90^{0}(\uparrow)$ and $%
135^{0}(\nwarrow)$. The bits of Alice are encoded from these
directions in the following way: $0^{0}$ and $45^{0}$ represent bit $0$, $%
90^{0}$ and $135^{0}$ represent bit $1$. When a photon arrives at Bob's
observation station, Bob performs a measurement on this photon based on
a randomly selected basis, either the rectilinear or the diagonal basis. Bob
encodes the outcome of his measurements in the same way as Alice does. If Bob
chooses a basis which is consistent with Alice selection (for instance,
Alice's sends a photon polarized in $0^{0}$ or $90^{0}$, and Bob performs the
measurement on the rectilinear basis), then it is assumed
that Bob's bit is identical to that of Alice.
Otherwise Bob will guess the wrong bit for about $50\%$\ of the detected photons.
Table~\ref{table1} lists the various possibilities that Alice and Bob may encounter
during the exchange of data.
It should be clear from this discussion that up to this point,
it has been assumed that the detectors operate with 100\% detection efficiency,
that the coordinate systems of Alice and Bob are perfectly aligned and so on,
that is we assume that the experiment is perfect.

After recording a collection of events, in the next step,
Alice and Bob will sift the key from the original raw bits by
communicating through a conventional classical channel. For each photon
that Bob has received, he tells Alice which basis he has selected but
he does not tell her the result of the measurement.
Then, for each photon, Alice announces to Bob whether he made a correct choice.
Finally they discard all the bits for which Bob has made the wrong choice of basis.
The bits that survive the sifting procedure constitute the key
and be used to encrypt the data that they want to send to each other.

An eavesdropper, conventionally called Eve, who attempts to intercept some
photons during the key transmission process will cause some
errors in the sifted key, and these errors can be detected by Alice and Bob
through publicly comparing randomly selected subsets of their sifted key.
If Eve performs the similar measurements as Bob on all photons sent by Alice,
and then prepares and resents new photons according to her measurements,
Alice and Bob will observe an error rate of about $25\%$ and conclude
that their communication channel is not secure.

\begin{table*}[tp]
\caption{The quantum theoretical predictions for the single- and two particle probabilities
of a system in the singlet state and in the product state.
The upper part shows the probability of observing $+$ on one side
and the joined probability of observing $+$ on both sides.
The lower part gives the expressions for the Wigner parameter $S$ and the modified Wigner parameter $S^\prime$.}
\label{table0}
\begin{center}
\begin{ruledtabular}
\begin{tabular}
[c]{ccc}
& Singlet state & Product state\\
\hline
${P}_{+}{(\phi}_{A}{)}$ & $\frac{1}{2}$ & $\cos
^{2}{(\psi}_{A}{-\phi}_{A}{)}$\\
${P}_{+}{(\phi}_{B}{)}$ & $\frac{1}{2}$ & $\cos
^{2}{(\psi}_{B}{-\phi}_{B}{)}$\\
${P}_{++}{(\phi}_{A}{,\phi}_{B}{)}$ & $\frac{1}%
{2}\sin^{2}{(\phi}_{A}{-\phi}_{B}{)}$ & $\cos
^{2}{(\psi}_{A}{-\phi}_{A}{)}\cos^{2}{(\psi}%
_{B}{-\phi}_{B}{)}$\\\hline
${S}$ & $\sin^{2}{\theta-}\frac{1}{2}\sin^{2}{2}%
{\theta}$ & $%
\begin{array}
[c]{c}%
\cos^{2}{\psi}_{A}\cos^{2}{(\psi}_{B}{-\theta)+}\cos
^{2}{(\psi}_{A}{+\theta)}\cos^{2}{\psi}_{B}-\\
\cos^{2}{(\psi}_{A}{+\theta)}\cos^{2}{(\psi}%
_{B}{-\theta)}%
\end{array}
$\\
${S}^{\prime}$ & $\sin^{2}{\theta-}\frac{1}{2}\sin^{2}%
{2}{\theta}$ & $%
\begin{array}
[c]{c}%
\cos^{2}{\psi}_{A}\cos^{2}{(\psi}_{B}{-\theta)+}\cos
^{2}{(\psi}_{A}{+\theta)}\cos^{2}{\psi}_{B}{+}\\
{\sin}^{2}{\psi}_{A}\sin^{2}{\psi}_{B}{-}
\cos^{2}{(\psi}_{A}{+\theta)}\cos^{2}{(\psi}_{B}%
{-\theta)}%
\end{array}
$\\
\end{tabular}
\end{ruledtabular}
\end{center}
\end{table*}

\section{Ekert's protocol}
\label{EKERT}

In 1991, Artur Ekert proposed another protocol based on entanglement states
with security guaranteed by Bell inequalities~\cite{Ekertref}.
The source can be any two-particle system with some property entangled.
In the original proposal of the protocol,
pairs of spin-$\frac{1}{2}$ particles in a singlet state are used as the information carrier,
but in the real experiments~\cite{bellref1,bellref2},
polarization entangled photon pairs are most commonly used to implement this protocol.
In Ref.~\cite{bellref2}, the CHSH inequality (one of the many forms of a Bell inequality)
is used to test of the security.
In Ref.~\cite{bellref1} another form of Bell inequality, the Wigner inequality~\cite{wigref},
provides a relative simple test of the security.
In this section, we briefly review the main idea of this protocol.

First, it is assumed that there is a source that emits
pairs of photons, one photon traveling to Alice
and the other photon traveling to Bob.
It is assumed that the state of the whole system, that is
the description of the observation of the polarization of many pairs,
can be described by the singlet state

\begin{equation}
|\Psi \rangle =\frac{1}{\sqrt{2}}(|H\rangle _{A}|V\rangle _{B}-|V\rangle
_{A}|H\rangle _{B}),
\label{singlet}
\end{equation}%
where $H$ and $V$ denote the horizontal and vertical (linear) polarization
states.
When a pair of photons $A$ and $B$ has been generated at the source,
they are spatially separated and sent to Alice and Bob
through free air or through some special optical fiber.
Then Alice and Bob perform measurements on the
polarization state of the received photon using a polarizing beamsplitter.
Both Alice and Bob independently and randomly select between two polarization
orientations.
Let us denote the two orientations of Alice by
$\phi _{A_{1}}$ and $\phi _{A_{2}}$, and those of Bob by
$\phi _{B_{1}}$ and $\phi _{B_{2}}$.
The outcome for an individual measurement is represented by either $+1$ or $-1$.
Because of the assumed entanglement between the polarization of the two photons,
if Alice and Bob select parallel but otherwise arbitrary orientations of their polarizer,
the outcomes of these two measurements are expected to display perfect anticorrelation.
Thus, anticorrelation between the two measurements with $\phi _{A_{1}}=\phi _{B_{1}}$  can be used to establish the key.

When the two photons of a particular pair are measured in two non-parallel orientations,
the correlation between them cannot be recognized as such.
Then, we need a test to see if the correlation are those of a system in the singlet state.
The Wigner inequality provides a convenient tool to do this.
We denote by $P_{++}(\phi_{A_{1}},\phi_{B_{2}})$, $P_{++}(\phi_{A_{2}},\phi_{B_{1}})$, and
$P_{++}(\phi_{A_{2}},\phi_{B_{2}})$ the probabilities to obtain $+1$ on
both sides for these three pairs of different orientations of the polarizers.
Under the assumptions discussed in Appendix~\ref{AppendixA}
these three probabilities must obey Wigner inequality
\begin{equation}
P_{++}(\phi _{A,1},\phi _{B,2})+P_{++}(\phi _{A,2},\phi _{B,1})-P_{++}(\phi
_{A,2},\phi _{B,2})\geq 0.
\label{Wigner inequality}
\end{equation}%
In Appendix~\ref{AppendixA}, we give a simple proof of the Wigner inequality.
For later use, it is expedient to define the Wigner parameter by
\begin{equation}
S=P_{++}(\phi _{A,1},\phi _{B,2})+P_{++}(\phi _{A,2},\phi _{B,1})-P_{++}(\phi_{A,2},\phi _{B,2})
\label{WignerS}
\end{equation}%

For the singlet state Eq.~(\ref{singlet}), quantum theory predicts that
\begin{equation}
P_{++}(\phi _{A,1},\phi _{B,1})=\frac{1}{2}\sin ^{2}(\phi _{A,1}-\phi_{B,1}).
\label{QMprediction1}
\end{equation}%
Inserting Eq.~(\ref{QMprediction1}) into Eq.~(\ref{Wigner inequality}),
it is easy to check (see later for examples) that for some a range of
$\phi_{A,1}$, $\phi_{B,2}$, $\phi_{A,2}$, and $\phi_{B,2}$,
we have $S<0$. Hence the Wigner inequality Eq.~(\ref{Wigner inequality}) is
violated by a quantum system in the singlet state.
For a particular choice of orientations that is used to implement
Ekert's protocol namely $\phi _{A,1}=\phi _{B,1}=0$, $\phi _{A,2}=30^{0}$,
and $\phi _{B,2}=-30^{0}$, we find that $S=-1/8$.
Any attempt to tamper with the singlet state will
change $S$ from its minimum value $-1/8$ to a larger one.

What happens to the observed data when photons are intercepted and resend?
Is the Wiger inequality still powerful enough to reveal the insecurity of the whole system?
It turns out that the assumption of perfect anticorrelation, essential for the
derivation of Wigner inequality, may cause a security problem.
To alleviate this problem, a modified Wigner inequality that does not rely
on the assumption of perfect anticorrelation was introduced in Ref.~\cite{modwigref}.
An experimental test of the power of the modified Wigner inequality
in the presence of eavesdropping is given in Ref.~\cite{wigqcref}.
A simple proof of this Modified Wigner inequality is
given in Appendix~\ref{AppendixB}.
Introducing the modified Wigner parameter $S^\prime$ by
\begin{eqnarray}
S^{\prime }&=&P_{++}(\phi _{A,1},\phi _{B,2}) +P_{++}(\phi _{A,2},\phi_{B,1})
\notag \\
&+&P_{--}(\phi _{A,1},\phi _{B,1}) -P_{++}(\phi _{A,2},\phi _{B,2}),
\end{eqnarray}
the modified Wigner inequality reads
\begin{eqnarray}
S^{\prime }\ge 0.
\end{eqnarray}
Compared to the original Wigner inequality,
an extra term is added which contributes when
both Alice and Bob choose the same orientation of their
polarizers. This extra term significantly increases
the possibility of detecting the presence of an eavesdropper.

For the simulation of this protocol in the presence of eavesdropping, we
assume that Eve performs the intercept-resend strategy.
This implies that Eve detects
the two photons using two polarizers with orientations $\psi_{A}$ and $\psi_{B}$,
respectively, and then uses the result of her measurement of the two photons
to prepare a photon that she sends to Alice and another photon
that she sends to Bob.
According to quantum theory, the photons received by both
Alice and Bob are described by the product state
\begin{eqnarray}
|\Psi \rangle &=&(\cos \psi _{A}|H\rangle _{A}+\sin \psi _{A}|V\rangle _{A})
\notag \\
&&\times (\cos \psi _{B}|H\rangle _{B}-\sin \psi _{B}|V\rangle _{B}).
\end{eqnarray}%
The predictions of quantum theory for the
product state and the singlet state are summarized in Table~\ref{table0}.

In Table~\ref{table0}, ${\phi }_{A}$ and ${\phi }_{B}$ denote the
orientations of Alice's and Bob's polarizer, respectively,
${P}_{+}{(\phi }_{A}{)}$ represents the probability that Alice obtains
$+1$ in her measurement, ${P}_{+}{(\phi }_{B}{)}$ denotes the
corresponding probability for Bob's measurement,
and ${P}_{++}{(\phi }_{A}{,\phi }_{B}{)}$ refers to the probability that
both Alice and Bob record a $+1$ result.
In the last two rows of Table~\ref{table0}
we list the results of quantum theory for the original and the modified Wigner parameter
for the polarizer settings $\phi _{A,1}=\phi _{B,1}=0$, $\phi _{A,2}=-{\theta }$,
and $\phi _{B,2}={\theta }$.
For the singlet state, the additional term in the modified Wigner parameter is zero.
Hence there is no difference between $S$ and $S^{\prime }$
and both of them only depend on ${\theta }$.
However in the case of a product state,
the additional term ${\sin}^{2}{\psi}_{A}\sin^{2}{\psi}_{B}$ that
depends on the polarization states of the photons
and is always non-negative, hence $S^\prime\ge S$.
Thus, if Alice and Bob expect to observe the singlet state they should find
$S^\prime= S=-1/8$. However,
if Eve is intercepting and sending photons in a product state,
the modified Wigner inequality provides more
power to disclose the existence of an eavesdropper
because Eve's actions will cause $S^\prime$ to change
from being negative to positive while $S$ may remain negative
(see the examples shown later).

\begin{table*}[tp]
\caption{The first $100$ bits of the sifted key of the BB84 protocol without
eavesdropping.
The upper part gives bits 0 -- 50, the lower part gives bits 51 -- 100.
As expected, the sifted key of Alice is identical to the one obtained by Bob.}
\label{table2}
\begin{center}
\begin{ruledtabular}
\begin{tabular}
[c]{cccccccccccccccccccccccccccccccccccccccccccccccccccccc}

Alice's bits &1&0&0&0&0&0&0&1&1&1&0&0&1&1&1&0&0&1&1&0&1&1&0&0
&0&1&1&0&1&0&0&0&0&1&0&0&1&1&1&0&1&0&1&1&1&0&1&0&1&0\\
Bob's bits &1&0&0&0&0&0&0&1&1&1&0&0&1&1&1&0&0&1&1&0&1&1&0&0
&0&1&1&0&1&0&0&0&0&1&0&0&1&1&1&0&1&0&1&1&1&0&1&0&1&0\\\hline

Alice's bits &1&0&1&1&1&1&1&1&0&1&0&0&1&0&1&0&1&0&1&1&0&1&0&0
&1&0&1&1&1&1&1&0&1&1&1&0&1&1&1&0&0&1&0&1&0&0&0&0&1&0\\
Bob's bits &1&0&1&1&1&1&1&1&0&1&0&0&1&0&1&0&1&0&1&1&0&1&0&0
&1&0&1&1&1&1&1&0&1&1&1&0&1&1&1&0&0&1&0&1&0&0&0&0&1&0

\end{tabular}
\end{ruledtabular}
\end{center}
\end{table*}

\section{Event-based simulation of a polarizer}
\label{POL}

Both the practical realization of the
BB84 and the Ekert protocol use the detection of the photon polarization.
Hence, the polarizer is an indispensable apparatus for both Alice and Bob to perform
their measurements.
Therefore, to set up an event-by-event computer simulation model
for these quantum cryptography protocols,
we first need to consider event-based simulation models for a polarizer.

Some optically active materials such as calcite
split an incoming beam of light into two spatially
separated beams depending on the polarization property of the incident beam~\cite{BORN64}.
If the incident beam has polarization $\psi $
and the orientation of the polarizer is denoted by $\phi $,
the intensities of the two output beams $0$ and $1$
are given by Malus' law
\begin{align}
I_{0}& =\cos ^{2}(\psi -\phi ),  \notag \\
I_{1}& =\sin ^{2}(\psi -\phi ).  \label{Malus law}
\end{align}%
The polarization of output beam $0$ is $\phi $, and the polarization of output beam $1$
is $\phi +\pi /2$.
The incident beam is said to be randomly polarized if $I_{0}=I_{1}=1/2$.

The simplest simulation model of a polarizer
determines the type ($0$ or $1$) of the output
by comparing a uniform pseudo-random number $r$ with $\cos ^{2}(\psi -\phi )$.
If $r\leq \cos ^{2}(\psi -\phi)$, the output is of type $0$, otherwise it is of type $1$.
For each individual input photon, the outcome is pseudo-random,
but if we repeat this process for sufficiently many events
(and the pseudo-random number generator is of sufficient quality),
the frequencies of observing photons in output $0$ and $1$
will agree with Malus' law. In the sequel, this model
for the polarizer will be called the probabilistic polarizer (PP).

As an alternative for the PP, we will also simulate
both protocols using a deterministic model
for the polarizer~\cite{RAED05b,RAED05c,RAED05d,RAED06a,MICH05}.
This model will be called the deterministic polarizer (DP).
As the details of this model are not of importance for
(the analysis of) the results presented in this paper,
we refer the reader who is interested in this and other
deterministic simulation models for quantum phenomena
to Refs.~\onlinecite{RAED05b,RAED05c,RAED05d,RAED06a,MICH05}.
In this paper, we show that both the PP and DP models are capable
of reproducing {\it exactly} all the results of quantum
theory for both the BB84 and the Ekert protocol
using an event-by-event based simulation algorithm.

\begin{table*}[tp]
\caption{The first $100$ bits of the sifted key of the BB84 protocol in the
presence of eavesdropping.
The upper part gives bits 0 -- 50, the lower part gives bits 51 -- 100.
The differences between Alices's and Bob's sifted key are emphasized by underlining the bits.
The error rate is about 26\%.}
\label{table3}
\begin{center}
\begin{ruledtabular}
\begin{tabular}
[c]{cccccccccccccccccccccccccccccccccccccccccccccccccccc}

Alice's bits&$\underline{1}$&0&1&0&0&0&0&$\underline{1}$&0&0&
1&$\underline{1}$&1&1&0&0&0&0&0&0&
$\underline{1}$&0&1&0&0&0&0&0&1&$\underline{1}$&
$\underline{1}$&1&0&1&$\underline{1}$&1&0&1&1&0&
$\underline{1}$&1&1&0&0&1&1&0&1&1\\ 

Bob's bits&$\underline{0}$&0&1&0&0&0&0&$\underline{0}$&0&0&
1&$\underline{0}$&1&1&0&0&0&0&0&0&
$\underline{0}$&0&1&0&0&0&0&0&1&$\underline{0}$&
$\underline{0}$&1&0&1&$\underline{0}$&1&0&1&1&0&
$\underline{0}$&1&1&0&0&1&1&0&1&1 \\ \hline

Alice's bits&$\underline{1}$ &0&1&1&$\underline{1}$&1&0&1&$\underline{0}$&$\underline{1}$&0&0&$\underline{0}$&$\underline{0}$&
$\underline{0}$&0&1&0&$\underline{0}$&1&$\underline{1}$&1&0&0&1&0&1&1&0&0&1&
$\underline{1}$&0&$\underline{0}$&1&0&1&1&1&0&$\underline{0}$&$\underline{0}$&$\underline{1}$&
$\underline{1}$&$\underline{0}$&$\underline{1}$&1&1&$\underline{0}$&0 \\ 

Bob's bits&$\underline{0}$&0&1&1&$\underline{0}$&1&0&1&$\underline{1}$&$\underline{0}$&0&0&$\underline{1}$&$\underline{1}$&
$\underline{1}$&0&1&0&$\underline{1}$&1&$\underline{0}$&1&0&0&1&0&1&1&0&0&1&
$\underline{0}$&0&$\underline{1}$&1&0&1&1&1&0&$\underline{1}$&$\underline{1}$&$\underline{0}$&
$\underline{0}$&$\underline{1}$&$\underline{0}$&1&1&$\underline{1}$&0

\end{tabular}
\end{ruledtabular}
\end{center}
\end{table*}

\section{Event-based simulation of the BB84 protocol}
\label{BB84SIM}

In this section, we present the results of an event-by-event simulation
of the BB84 protocol using both PPs and DPs.
We start with the original BB84 protocol and demonstrate that
the sifted bits obtained by Alice and Bob are identical, as expected.
Then, we simulate the effect of eavesdropping by Eve who uses the intercept-resend strategy
and show that this introduces significant errors in the sifted key.
Finally, we study the effect of misalignment of the settings of both Alice's and Bob's polarizers
by computing the fidelity of the sifted key as a function of the misalignment angles.

\subsection{Simulation of the BB84 protocol in the absence of an eavesdropper}

Two polarizers are needed to simulate this protocol, one for Alice, and another one for Bob.
Alice uses her polarizer to encode the bits she wants to send
by (randomly) selecting the polarization $0^\circ(\longrightarrow),45^\circ(\nearrow )
,90^\circ(\uparrow )$ or $135^\circ(\nwarrow )$.
Thus, if Alice wants to encode a random sequence of bits,
Alice's polarizer $\phi _{A}$ chooses randomly from these four directions.
In the simulation, we use uniform pseudo-random numbers to select the polarizations,
implying that half of the photons leave the polarizer through output channel $0$ while the
others leave the polarizer through output channel $1$.
The photons leaving through channel $1$ are discarded and
all the other photons are sent to Bob.
The orientation $\phi _{B}$ of Bob's polarizer switches randomly between
$0^{0}(\longrightarrow )$ and $45^{0}(\nearrow )$. These two directions
define Bob's two measurement basis, the rectilinear or diagonal linear basis.

As indicated in Table~\ref{table1}, the probabilities to observe a photon
in output channel $0$ and $1$ are $\cos ^{2}(\phi_{A}-\phi _{B})$ and $\sin ^{2}(\phi _{A}-\phi _{B})$,
respectively.
It is clear from Table \ref{table1} that the cases for which Bob has a definite
outcome correspond to situation in which Bob selected an observation basis
that is consistent with the choice made by Alice.

In Table \ref{table2}, we show the first 100 bits of the sifted sequence, extracted from
a simulation sequence of $10^{5}$ events.
After passing through Alice's polarizer, the number of the photons sent to Bob is $49920$
(approximately $1/2$ of the total number of events), and the length
of the sifted key is $25072$ (approximately $1/4$ of the total number of
events). From Table \ref{table2}, we see that the first 100 bits of the
sifted key that Alice and Bob obtain are identical, as expected.
We have checked that the other bits in the sifted key are identical also (results not shown).
Furthermore, the simulation results for the sifted key do not depend the choice of the simulation
model (PP or DP) for the polarizer (results not shown).
The fidelity $F$, defined as the ratio of the correct bits in the sifted key
to the length of the sifted key is $100\%$.

\subsection{Simulation of the BB84 protocol in the presence of an
eavesdropper}

The presence of Eve is built into the simulation algorithm
by adding a polarizer and implementing the intercept-resend strategy.
Thereby, we assume that Eve intercepts all the photons that Alice
sends to Bob and that she is able to perform the measurements in the same
rectilinear or diagonal linear basis as the ones used by Bob.
It is easy to see that for this type of eavesdropping,
the error rate in the sifted key should be about $25\%$.
In Table \ref{table3}, we show the first 100 bits of the sifted sequence, extracted from
a simulation sequence of $10^{5}$ events.
The number of the photons that Bob receives
is $50074$ (approximately $1/2$ of the total number of events),
and the length of the sifted key is $25043$ (approximately $1/4$ of the total number of events).
In this case, the fidelity is about $75.2\%$.
As in the case without eavesdropper, the simulation results for the sifted key do not depend on the choice of the simulation
model (PP or DP) for the polarizer (results not shown).

\begin{center}
\begin{figure}[tp]
\begin{center}
\includegraphics[width=8cm]{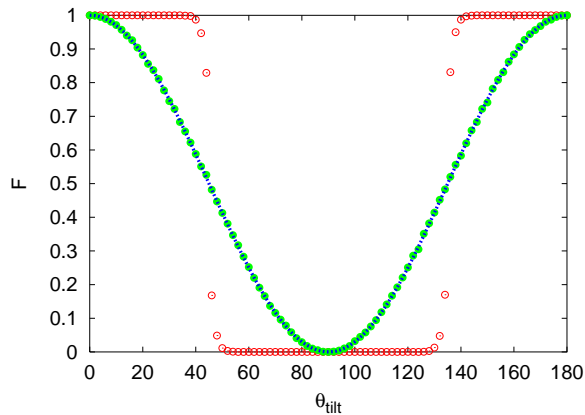}
\end{center}
\caption{The fidelity $F$ of the sifted key as a function of misalignment $\theta_{tilt}$.
Solid circles (green): Simulation data using the PP model.
Open circles (red): Simulation data using the DP model.
Dashed line (blue): Theoretical prediction given by Eq.~(\ref{fidelityeq}).}
\label{fidelity}
\end{figure}
\end{center}

\subsection{Misalignment of the measurement basis}

In this section, we consider a situation in which there is a misalignment of
the measurement basis of both parties.

In real experiments, it is unlikely that, say
the rectilinear basis used by Alice is perfectly aligned with
the rectilinear basis used by Bob.
Furthermore, in real experiments the polarization of the photons
changes as they propagate through the medium because of interactions
with the medium (air, fibers).
Moreover, for some strategies, the results of eavesdropping
can be viewed as a rotation of Bob's measurement basis, making
it more difficult to distinguish between a real misalignment
of the basis and the presence of an eavesdropper.
Thus, it is important to study the effects of
misalignments of the coordinate systems.

In our simulations, we use the orientations of Alice's polarizer as reference
and tilt Bob's basis by an angle $\theta _{tilt}$.
Therefore Bob's basis changes to from $0^\circ$ to $0^\circ+\theta _{tilt}$ and
from $45^\circ$ to $45^\circ+\theta _{tilt}$.
In Fig.~\ref{fidelity}, we show our simulation results for the
fidelity as a function of $\theta_{tilt}$.

Unlike in the two previous cases, the simulation results
for the sifted key depend on the choice of the simulation
model (PP or DP) for the polarizer.
For the PP model, the simulation data (solid circles in Fig.~\ref{fidelity})
is in good agreement with experimental results~\cite{ftqc2,ftqc1}.
For this model, we may compute the averaged fidelity by averaging (according to Malus' law)
the probabilities for obtaining identical bits in the sifted keys. We find
\begin{equation}
F_{PP}=\cos ^{2}\theta _{tilt}.  \label{fidelityeq}
\end{equation}
This function is shown as the dashed line in Fig.~\ref{fidelity},
demonstrating that there is excellent agreement between theory and simulation.
The open circles in Fig.~\ref{fidelity} are the simulation data as obtained with the DP model.
Clearly, the sifted keys obtained by using the DP model are much more robust
with respect to misalignments of the polarizers than the ones obtained by using the PP model.
On the other hand, the DP results do not agree with currently available
experimental results~\cite{ftqc2,ftqc1}, suggesting that the
polarizers that are used in these experiments are not
described by the DP model.

\subsection{Discussion}
\label{BB84SIMD}

Clearly, our simulation algorithm for the BB84 does not solve an equation
of quantum theory nor does it rely on concepts of quantum theory.
This should not come as a surprise: As quantum theory does not describe individual events (the
quantum measurement paradox)~\cite{HOME97,BALL03},
there is no reason to expect that quantum theory has any bearing
on quantum cryptography, other than that it describes the averages over many events.
As this point of view is in conflict with
popular statements that quantum cryptography requires
a full quantum mechanical description~\cite{QCintroduction} or
that quantum cryptography relies on the Heisenberg uncertainty relation,
it is of interest to consider the question at which point
concepts of quantum physics enter into our event-by-event simulation of the BB84 protocol.

From the description of the simulation algorithm, it is clear that
in order for the BB84 to be secure, it is essential
that the message and messenger have the following properties:
\begin{enumerate}
\item{A message can be one out of two pairs of possible items only.}
\item{The messenger tells the recipient which of the two pairs the item
that the messenger carries belongs to, but the messenger cannot tell a recipient which item it is.}
\item{The messenger can deliver the message only once and after delivering the message, the messenger self-destructs.}
\end{enumerate}

It is not too difficult to build a macroscopic device with these properties (minor modifications to
intelligent containers used for transport of valuables and cash would do).
Imagine that Alice has a set of boxes (the messengers) with the following properties:
\begin{enumerate}
\item{Once closed, the box explodes when it is being tampered with.
The box is shielded such that when it detects penetrating radiation, it explodes, making it impossible
to analyze its content without destroying the content.
Note that for secure quantum cryptography, similar conditions apply to Alice's and Bob's station too~\cite{QCintroduction}.}
\item{As long as the box is open, Alice can wire the electronics inside the box such that the electronic circuit
encodes one of the four possibilities according to Table ~\ref{table1}.
After wiring her bit, she closes the box and sends it to Bob.}
\item{On the outside, the box has a button and a switch, the setting of which corresponds to
Bob's choice of the orientation of his measurements basis (see Table ~\ref{table1}).
Bob puts the switch in one of its two positions and then he presses the button.
The electronics inside the box, causes the box to explode immediately, after five seconds
or after ten seconds, corresponding to the case where Bob detects a $0$, ?, or 1, respectively (see Table ~\ref{table1}).
}
\end{enumerate}
It is not difficult to see that this classical, macroscopic device is no less vulnerable to eavesdroppers than
the quantum cryptography system. Of course, the latter is much more user-friendly and less expensive to operate.

\section{Event-based simulation of the Ekert protocol}
\label{EKERTSIM}

Starting from the observation
that coincidence in time is a key ingredient in experimental realizations of the EPR \textit{gedanken} experiment,
several computer simulation algorithms have been proposed that (1)
satisfy Einstein's conditions of local causality and realism
and (2) exactly reproduce the two-particle correlation that is characteristic for a quantum system
in the singlet state~\cite{RAED06c,RAED07b,ZHAO07a,RAED07c}.
These algorithms generate the data event-by-event,
use integer arithmetic and elementary mathematics to analyze the data,
and do not rely on concepts of probability theory or quantum theory.

In this Section, we use these algorithms to perform an even-by-event simulation of Ekert's quantum cryptography protocol.
For the sake of brevity, we do not review all the details of the algorithms. The reader who is interested in these aspects
should consult the original papers~\cite{RAED06c,RAED07b,ZHAO07a,RAED07c}.
The rigorous, probabilistic treatment given in Appendix~\ref{AppendixC}
proves that our classical simulation model reproduces the correlations that are characteristic
for a quantum system in the singlet state.

\subsection{Simulation of the Ekert protocol in the absence of an eavesdropper}

\begin{figure*}[tp]
\begin{center}
\includegraphics[width=18cm]{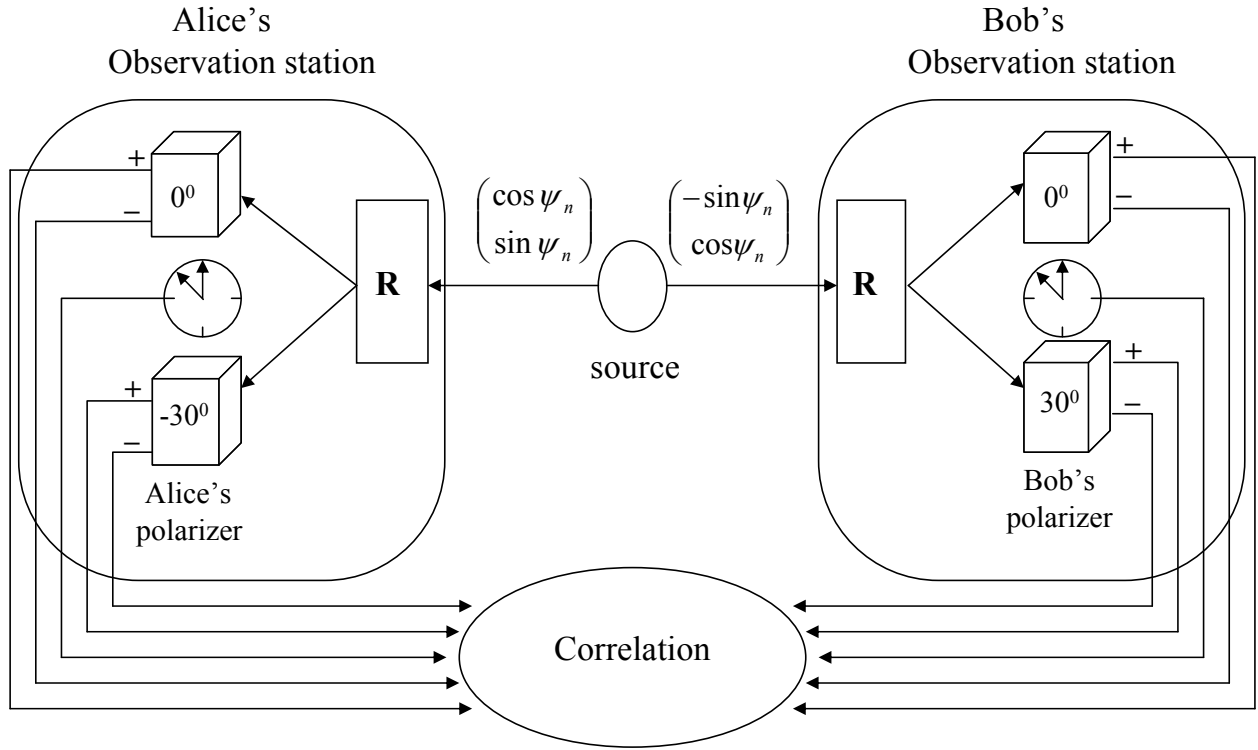}
\end{center}
\caption{Schematic diagram of the event-by-event simulation of Ekert's quantum cryptography protocol,
using the (modified) Wigner inequality as a guard against eavesdroppers.
The source emits pairs of particles with orthogonal but random polarization.
The particles fly to Alice's and Bob's observation station, respectively.
The rectangular boxes labeled by $R$ direct the particles to one of the two
polarizers, using a binary pseudo-random number.
When the particle emerges from a polarizer it generates
a $+1$ or $-1$ event, and a clock is used to attach a time tag to this event.
After collecting all events, Alice and Bob use time coincidence to correlate
their data and to extract the key.}
\label{simueprqc}
\end{figure*}

A schematic diagram of the simulation procedure is shown in Fig. \ref{simueprqc}.
In the simulation algorithm, the source generates pairs of particles (photons in the real experiment).
Particle A and B travel to Alice and Bob, respectively.
Each particle carries a two-dimensional unit vector given by
\begin{align}
S_{n,1} & =(\cos\psi_{n},\sin\psi_{n}),  \notag \\
S_{n,2} & =(-\sin\psi_{n},\cos\psi_{n}).
\end{align}
where $n$ labels the number of the event.
The vectors $S_{n,1}$ and $S_{n,2}$ represent the polarizations $\psi _{n}$ and $\psi _{n}+%
\frac{\pi }{2}$ of the two photons that fly to Alice and Bob, respectively.
The distribution of $\psi _{n}$ is taken to be uniform over the interval $[0,2\pi ]$.
Note that after projecting the vectors $S_{n,1}$ and $S_{n,2}$ onto
dichotomic variables, the latter satisfy the conditions for deriving
the Wigner inequality (see Appendix~\ref{AppendixA}).

When the particle arrives at Alice's (Bob's) station, labeled by $i=1$ ($i=2$),
a random number is used to select the polarizer
that will be used to perform the polarization measurement on the photon.
This measurement maps the angle $\psi_{n}$ onto the variable $x_{n,i}=x_{n,i}(\psi _{n},A_{n,i})=\pm1$.
Thus, the results of generating $N$ of these events can be summarized
by
\begin{eqnarray}
\Gamma _{n,i} &=&\{x_{n,i}=\pm 1, A_{n,i}=\pm1|n=1,...,N\},
\end{eqnarray}
where $A_{n,i}$ denotes which of the polarizers has been selected.
It is clear that we have assumed that the value of $x_{n,i}$ depends on
the incoming polarization and the internal orientation of the selected
polarizer only.

In any real experiment, one needs a criterion to decide whether
two objects form a single two-particle system or whether
they may be considered as two single-particle systems.
EPR experiments are no exception to this~\cite{WEIH98,WEIH00}.
EPR experiments with photons use coincidence in time to identify a single pair of two photons.
Note that time coincidences play an essential role
in real quantum cryptography experiments~\cite{QCintroduction}.

In practice, Alice and Bob add time tags to their detection events
in order to be able to count coincidences.
As the optical components (polarizers) induce time delays,
it is reasonable for a particle to experience a time delay when it
passes through the detection system.
To mimic this, we introduce the time delay into our simulation algorithm~\cite{RAED06c,RAED07b,ZHAO07a,RAED07c}.
At each station, we generate a time tag that depends on the local settings only.
Then, we compare the difference between the two time tags with a certain time window $W$.
If this difference is smaller than $W$, the detection events are considered to be coincident.
Otherwise, they are discarded.

We assume that the maximum time delay $T_{n,i}$
for a particle passing through a polarizer depends only on the angle
difference between the polarization of the incident particle and the
internal orientation of the polarizer. For instance, on Alice's side,
we set $T_{n,1}=T_{n,1}(\psi _{n}-\phi _{A,i})$.
The time tag $t_{n,i}$ itself is taken to be a pseudo-random
number from the interval $[0,T_{n,1}]$~\cite{RAED06c,RAED07b,ZHAO07a,RAED07c}.
Summarizing, the simulation algorithm generates two data sets
\begin{eqnarray}
\Upsilon_{n,i} =\{x_{n,i}=\pm1,A_{n,i}=\pm1,t_{n,i} |n=1,...,N\},
\label{Collecteddata}
\end{eqnarray}
for $i=1$ (Alice) and $i=2$ (Bob).
The structure of these data sets is identical to the data sets collected in EPR experiments with
photons~\cite{WEIH98,WEIH00}.

From Ref.~\cite{RAED06c,RAED07b,ZHAO07a,RAED07c}, we know that
the simulation model can reproduce all the results of quantum theory
of a system of two $S=1/2$ particles if we take
$T_{n,1}(\theta)=\left\vert \sin 2\theta\right\vert ^{d}$
(note that we have chosen the maximum time delay as the unit of time).
Here $d$ is a free parameter, which we call the time-delay parameter.
If $d=0$, we have $T_{n,i}=1$, implying that the
maximum time delay does not depend on the relative orientation.
In this case, the time delay has no essential influence on the final results~\cite{RAED06c,RAED07b,ZHAO07a,RAED07c}.
In our simulation (and also in experiment~\cite{WEIH98}),
we first fix the time-tag resolution, denoted by $0<\tau <1$.
Then, in our simulations, the time window is defined by $W=k\tau $, where $k$ is an integer.
It is clear that $\tau $ effectively determines the resolution by which we can resolve differences in the angles.
After generating $N$ pairs and collecting the data Eq.~(\ref{Collecteddata}),
we count the coincidences and we obtain an estimate for the probability
\begin{equation}
P_{++}(\phi _{A},\phi _{B})=\frac{C_{++}}{C_{++}+C_{--}+C_{+-}+C_{-+}},
\label{countp++}
\end{equation}%
where $C_{xy}\equiv C_{xy}(\phi _{A},\phi _{B})$
denotes the number of coincidences
between the signal $x =\pm $1 at station
1 and a signal $y =\pm 1$ at station 2
for a fixed combination of $\phi _{A}$ and $\phi _{B}$
and is given by
\begin{eqnarray}
\label{Cxy}
C_{xy}&=&\sum_{n=1}^N\delta_{x,x_{n ,1}} \delta_{y,x_{n ,2}}
\Theta(W-\vert t_{n,1} -t_{n ,2}\vert)
.
\end{eqnarray}
From Eq.~(\ref{countp++}) we compute the Wigner parameter $S$ according to Eq.~(\ref{WignerS}).

\begin{center}
\begin{figure}[tp]
\begin{center}
\includegraphics[width=8cm]{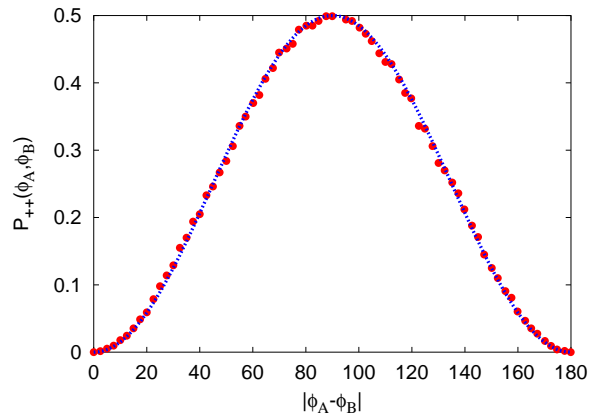}
\end{center}
\caption{%
$P_{++}(\phi _{A},\phi _{B})$ as a function of $\vert\phi _{A}-\phi _{B}\vert $.
Solid circles (red): Simulation data obtained by using the DP model with $d=2$.
Dashed line (blue): Quantum theory (Eq.~(\ref{QMprediction1})).
}
\label{figwigpp1}
\end{figure}
\end{center}

We first show simulation results obtained by using the DP model.
In Fig.~\ref{figwigpp1}, we plot the probability $P_{++}(\phi_{A},\phi_{B}) $
for fixed $\phi _{A}=0$ and $0\le\phi _{B}\le2\pi$.
The values of the other parameters used in the simulation are
$k=1$, $d=2$, $\tau =0.00025$, and $N=10^{8}$.
The dashed line in Fig.~\ref{figwigpp1} is the quantum theoretical prediction
Eq.~(\ref{QMprediction1}).
From Fig.~\ref{figwigpp1}, we conclude that there is an excellent agreement
between the simulation data and quantum theory.

\begin{figure}[tp]
\begin{center}
\includegraphics[width=8cm]{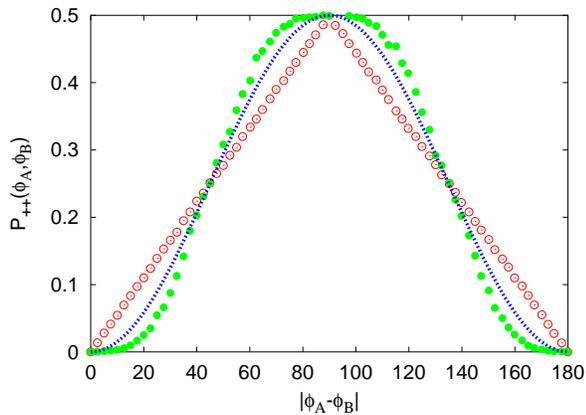}
\end{center}
\caption{%
$P_{++}(\phi _{A},\phi _{B})$ as a function of $\vert\phi _{A}-\phi _{B}\vert $.
Open circles (red): Simulation data by using the DP model with $d=0$.
Solid circles (green): Simulation data obtained by using the DP model with $d=4$.
Dashed line (blue): Quantum theory (Eq.~(\ref{QMprediction1})).
}
\label{figwigpp2}
\end{figure}

For comparison, in Fig.~\ref{figwigpp2} we present
the simulation results for $d=0$ and $d=4$.
For $d=0$, the two time tags that we generated are just two independent uniform pseudo-random numbers between $0$ and $1$
and contain no information about the polarizations of the incident photons or the orientations of the polarizers.
Therefore, because of the procedure to count coincidences,
the size of the window can only influence the numbers of events we collect: $W$ affects the statistical fluctuations only.
As a check, we have taken a fairly large time window ($W=100\tau$)
and found that in this case, the distribution $P_{++}(\phi_{A},\phi_{B})$ is
very close to the distribution that we find if we accept all the events (no coincidence window).
For $d=4$, we see from Fig.~\ref{figwigpp2} that the correlations are ``stronger'' than
those of the quantum system~\cite{RAED06c,RAED07b,ZHAO07a,RAED07c}.

Next, we consider the Wigner parameter $S$ for the fixed relation of the four
orientations mentioned above: $\phi_{A,1}=\phi_{B,1}=\varphi$, $\phi_{A,2}=\varphi-\theta,$ $\phi_{B,2}=\varphi+\theta$.
Inserting these special values into Eq.~(\ref{WignerS}) we get
\begin{equation}
S(\theta)=\sin^{2}\theta-\frac{1}{2}\sin^{2}2\theta.
\label{simplified WI}
\end{equation}
Since in this case, the Wigner parameter depends on $\theta$ only, it is sufficient to consider the case $\varphi=0$.

\begin{table*}[tp]
\caption{The first $100$ bits of the sifted key of Ekert's
protocol without eavesdropping, as obtained by using the DP model.
The upper part gives bits 0 -- 50, the lower part gives bits 51 -- 100.
In this case, there are no errors in this part of the sifted key (the error rate is very low ($10^{-5}$)).
}
\label{table4}
\begin{center}
\begin{ruledtabular}
\begin{tabular}
[c]{ccccccccccccccccccccccccccccccccccccccccccccccccccc}
Alice's bits &1&1&1&0&0&0&1&0&0&1&1&0&1&1&1&1&1&1&1&0&1&0&0&0&0&0&1
&0&0&0&0&1&0&1&1&1&0&1&0&0&1&1&1&1&0&1&0&1&0&1\\
Bob's bits &1&1&1&0&0&0&1&0&0&1&1&0&1&1&1&1&1&1&1&0&1&0&0&0&0&0&1
&0&0&0&0&1&0&1&1&1&0&1&0&0&1&1&1&1&0&1&0&1&0&1\\\hline
Alice's bits &1&0&0&1&0&1&1&0&0&1&0&1&0&0&1&0&0&0&1&1&0&1&0&0&1&0&1
&1&0&1&0&1&1&0&0&1&1&0&1&0&0&0&0&0&0&1&0&0&0&0\\
Bob's bits &1&0&0&1&0&1&1&0&0&1&0&1&0&0&1&0&0&0&1&1&0&1&0&0&1&0&1
&1&0&1&0&1&1&0&0&1&1&0&1&0&0&0&0&0&0&1&0&0&0&0\\
\end{tabular}
\end{ruledtabular}
\end{center}
\end{table*}

\begin{table*}[tp]
\caption{
The first $100$ bits of the sifted key of the Ekert protocol without eavesdropping,
as obtained by using the PP model.
The upper part gives bits 0 -- 50, the lower part gives bits 51 -- 100.
The differences between Alices's and Bob's sifted key are emphasized by underlining the bits.
There are only 2 errors in this part of the sequence.
}
\label{table5}
\begin{center}
\begin{ruledtabular}
\begin{tabular}
[c]{ccccccccccccccccccccccccccccccccccccccccccccccccccc}
Alice's bits&0&0&0&0&0&0&1&1&0&1&0&0&1&0&1&0&1&0&1&0&1&1&0&0&1&0&$\underline{0}$
&1&0&0&0&1&0&0&1&0&0&1&1&1&0&1&1&1&1&0&0&0&1&1\\
Bob's bits&0&0&0&0&0&0&1&1&0&1&0&0&1&0&1&0&1&0&1&0&1&1&0&0&1&0&$\underline{1}$
&1&0&0&0&1&0&0&1&0&0&1&1&1&0&1&1&1&1&0&0&0&1&1\\\hline
Alice's bits&1&1&1&0&0&0&1&0&1&1&0&0&0&1&0&1&1&$\underline{1}$&0&0&1&1&0&0&0&0&0
&0&0&1&0&0&1&0&0&1&0&1&1&0&1&0&0&0&1&0&0&0&0&1\\
Bob's bits&1&1&1&0&0&0&1&0&1&1&0&0&0&1&0&1&1&$\underline{0}$&0&0&1&1&0&0&0&0&0
&0&0&1&0&0&1&0&0&1&0&1&1&0&1&0&0&0&1&0&0&0&0&1\\
\end{tabular}
\end{ruledtabular}
\end{center}
\end{table*}

\begin{figure}[tp]
\begin{center}
\includegraphics[width=8cm]{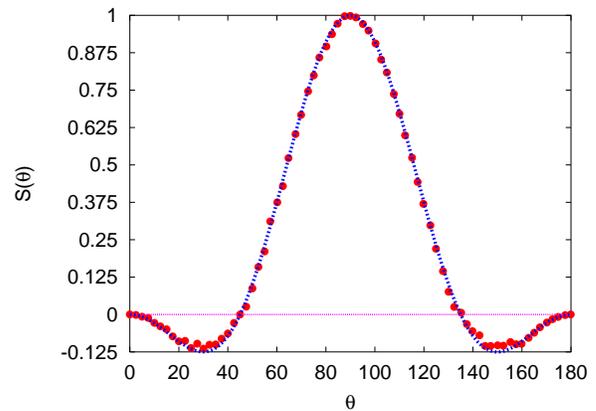}
\end{center}
\caption{%
The Wigner parameter $S$ as a function of $\theta $.
Solid circles (red): Simulation data obtained by using the DP model with $d=2$.
Dashed line (blue): Quantum theory (Eq.~(\ref{WignerS})).
}
\label{figwigs1}
\end{figure}

The simulation results are plotted in Fig.\ref{figwigs1}.
The values of the parameters used in the simulation are $k=1$,
$d=2$, $\tau =0.00025$, and $N=10^{8}$.
Again, we see an excellent agreement between the simulation data and quantum theory.
Furthermore, from Fig.\ref{figwigs1}
it is clear that the maximum violation of Wigner inequality is reached at $\theta =30^{0}$.
Therefore, in the Ekert protocol, the orientations of both parties are
chosen to be $\phi _{A,1}=\phi _{B,1}=0^{0}$, $\phi _{A,2}=30^{0}$, $\phi_{B,2}=-30^{0}$.
Then, the violation of the Wigner inequality signals the
strong anti-correlation of the pairs and the Wigner parameter $S$ can be used
to quantify the security of the protocol.

\begin{figure}[tp]
\begin{center}
\includegraphics[width=8cm]{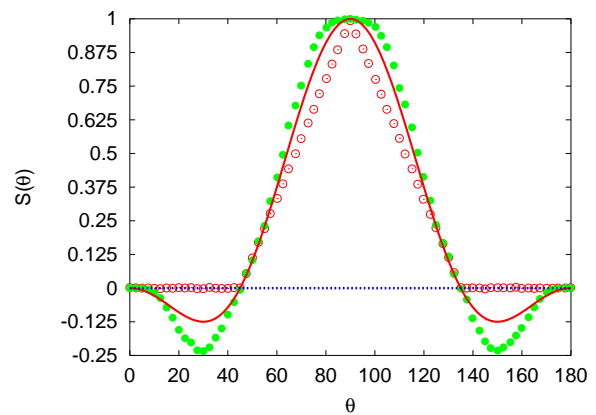}
\end{center}
\caption{
The Wigner parameter $S$ as a function of $\theta $.
Open circles (red): Simulation data obtained by using the DP model with $d=0$.
Solid circles (green): Simulation data obtained by using the DP model with $d=4$.
Solid line (red): Quantum theory (Eq.~(\ref{WignerS})).
}
\label{figwigs2}
\end{figure}

For completeness, we show in Fig.~\ref{figwigs2} the results
for the Wigner parameter for the cases $d=0$ and $d=4$.
As discussed above, $d=0$ corresponds to the case for which correlations are
computed without taking the time-tag information into account, showing
"classical" correlations.
For $d=4$, the correlation is stronger than the one of the quantum system,
hence the violation of the Wigner inequality can be larger.

Having established that our simulation algorithm reproduces the results
of quantum theory of a single system of two polarizations,
we now use the algorithm to simulate Ekert's quantum cryptography protocol.

As discussed earlier, the anticorrelated bits are generated using a
parallel basis (that is, the basis selected by both parties is
$\phi_{A,1}=\phi _{B,1}=0^{0}$).
After inverting all the bits from one of the two parties,
we expect to obtain two identical sequences.
The first $100$ bits from a long simulation are shown in Table~\ref{table4}.
In this simulation, which uses the DP model, we observe an almost perfect anticorrelation
of the two photons. Indeed, if $\phi _{A,1}=\phi _{B,1}=0^{0}$, the relative error
in the key is of the order of $10^{-5}$.

\begin{figure}[tp]
\begin{center}
\includegraphics[width=8cm]{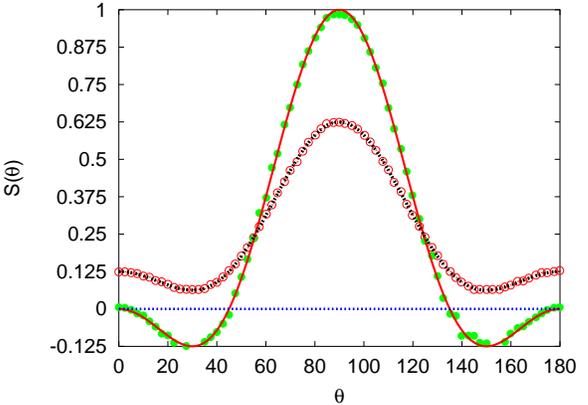}
\end{center}
\caption{The Wigner parameter $S$ as a function of $\theta $.
Solid circles (green): Simulation data obtained by using the PP model with $d=4$.
Open circles (red): Simulation data obtained by using the PP model with $d=0$.
Dashed line (blue): Quantum theoretical result for $S$ (see Table~\ref{table0});
Dashed line (black): Analytical results for $S$ and $d=0$ (see Appendix~\ref{AppendixC}).
}
\label{wignersimple}
\end{figure}

Finally, we simulate this protocol by using the PP model.
It is known that in order to reproduce the correct quantum correlations,
we must take $d=4$~\cite{ZHAO07a,RAED07c}.
Except for the value of $d$, we take the same simulation parameters
as in the DP-model simulations and repeat the calculation.
The simulation results are shown in Fig.~\ref{wignersimple}.

After inverting all the bits from one of the two parties,
we obtain two sequences of bit strings, the first $100$ bits
being shown in Table~\ref{table5}.
The error rate in this simulation is of the order of $10^{-2}$.
That this error rate is larger than in the DP simulation is easy
to understand: If we use the PP model, the outcome of each individual
measurement is inherently (pseudo-) random instead of deterministic as in the case of the DP model.

\subsection{Simulation of the Ekert protocol in the presence of an eavesdropper}

In the previous subsection, we have demonstrated that
by using the perfectly anticorrelated source,
together with the time-tag model and a time window to count coincidences,
we can reproduce the correlation that is characteristic for a quantum system in the singlet state.
In this subsection, we simulate the situation in which an eavesdropper is
present

First, we consider the special case in which Eve uses two polarizers with fixed,
perpendicular orientations: ${\psi }_{A}=45^{0}$ and ${\psi }_{B}=135^{0}$.
We should imagine that Eve can put these polarizers on both sides
of the source. Hence, she can manipulate the polarization
that Alice and Bob will observe in their measurements.

Our simulation model can easily deal with this complication: We just
put two PPs (or DPs) between the source and Alice and the source and
Bob, respectively. We repeat the simulations as in the case without an eavesdropper
and plot the two Wigner parameters $S$ and $S^\prime$ as a
function of $\theta $.

\begin{figure}[tp]
\begin{center}
\includegraphics[width=8cm]{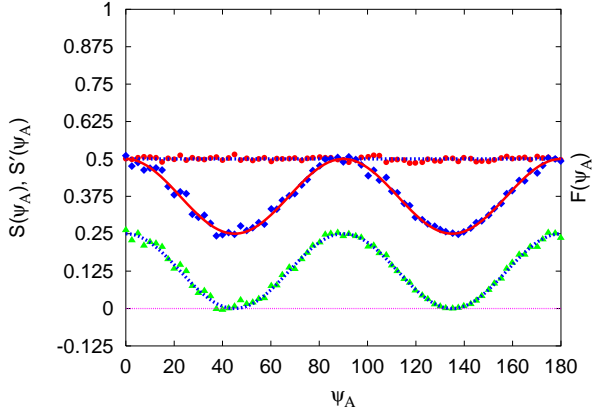}
\end{center}
\caption{The fidelity $F$ of the sifted key and the two Wigner parameters $S$ and $S^{\prime }$
as a function of $\theta$.
Solid circles (red): Simulation data for $F$;
Dashed line (blue): Quantum theoretical result for $F$ ($F=1/2$);
Solid diamonds (blue): Simulation data for $S^\prime$;
Solid line (red): Quantum theoretical result for $S^\prime$ (see Table~\ref{table0});
Solid triangles (green): Simulation data for $S$;
Dotted line (black): Quantum theoretical result for $S$ (see Table~\ref{table0}).}
\label{eve2}
\end{figure}

In Fig.~\ref{eve2}, we see two groups of curves with the same shape:
The simulation result of $S^{\prime }$ (blue solid diamonds)
and the quantum theoretical result of the modified Wigner parameter (red solid line)
agree very well.
The simulation result of $S$ (green solid triangles) and
the quantum theoretical result of the Wigner parameter (blue dashed line)
are in excellent agreement too.
Both the data for $S$ and $S^\prime$ are larger than zero,
signaling the presence of an eavesdropper.
As $S^\prime \ge S$, the modified Wigner parameter clearly is
more powerful to disclose the eavesdropper.

Also shown in Fig.~\ref{eve2}
is the fidelity of the sifted key as a function of $\theta$.
The simulation data for $F(\theta)$ lies on top of the theoretical expectation
$F(\theta)=1/2$.
It is clear that the fidelity does not depend on the angle $\theta $.
In this case, the value of the fidelity is about $0.5$
due to the choice of the orientations of Eve's polarizers.

\begin{figure}[tp]
\begin{center}
\includegraphics[width=8cm]{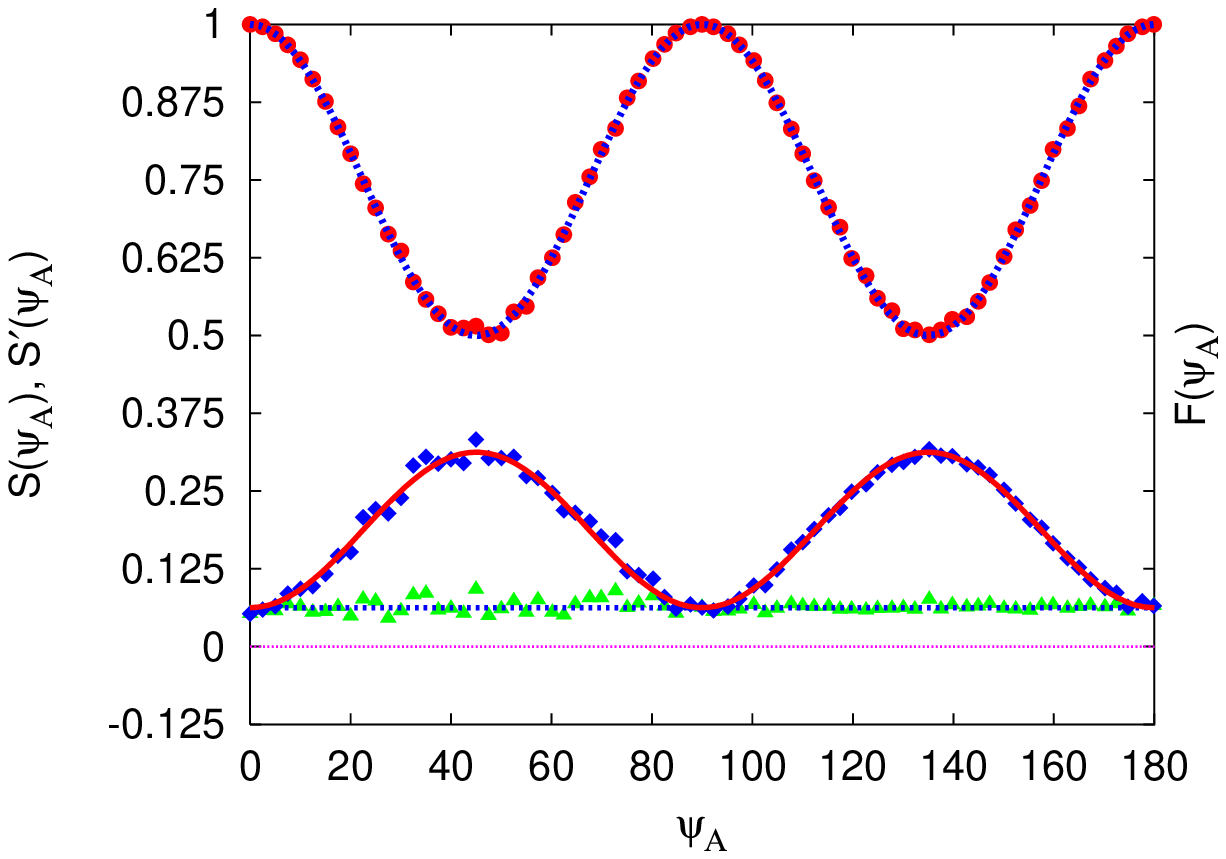}
\end{center}
\caption{The fidelity $F$ and the two Wigner parameter $S$ and $S^{\prime }$ as
a function of $\psi _{A}$ for the case ${\psi }_{B}=\psi _{A}+90^{0}$.
For the legend, see Fig.~\ref{eve2}.}
\label{eve1}
\end{figure}

\begin{figure}[tp]
\begin{center}
\includegraphics[width=8cm]{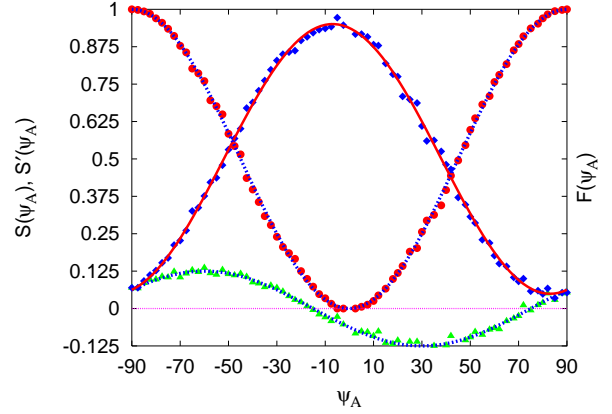}
\end{center}
\caption{The fidelity $F$ and the two Wigner parameter $S$ and $S^{\prime }$ as
a function of $\psi _{A}$ for the case ${\psi }_{B}=90^{0}$.
For the legend, see Fig.~\ref{eve2}.}
\label{eve3}
\end{figure}

Next, we take ${\theta }=30^{0}$ (the optimal values for Ekert's protocol
without eavesdropper) to study the dependence of the fidelity on the settings of Eve's polarizers.
We consider two different situations: The first one is that Eve's polarizers always
have perpendicular orientations.
The second one is that we fix one of Eve's polarizers at ${\psi }_{B}=90^{0}$,
and change the other setting ${\psi }_{A}$ gradually from $0^{0}$ to $180^{0}$.

The results for the first case are shown in Fig.~\ref{eve1}. The
upper curves show the dependence of the fidelity on the orientation of
Eve's polarizer: The solid circles represent the simulation data
and the dashed line is the theoretical result according to Malus' law
($F=1-\frac{1}{2}\sin ^{2}2{\psi }_{A}$).
The data in the middle are the simulation results (solid blue diamonds) and the
quantum prediction (red line) of the Wigner parameter $S^{\prime }$.
The bottom curves in Fig.~\ref{eve1} show the simulation data and the theoretical
result for the Wigner parameter $S$.
Recall that in order to detect the presence of an eavesdropper,
we must have $S>0$ or $S^\prime>0$ for all ${\psi }_{A}$.
Clearly, there is excellent agreement between theory and simulation.

The results for the second case are depicted in Fig.~\ref{eve3}.
The legend is the same as in Fig.~\ref{eve2}.
Again, there is excellent agreement between theory and simulation.
From Fig.~\ref{eve3}, it is clear that as $S<0$ for some range of angles,
using the Wigner parameter $S$ would not allow Alice and Bob to recognize
the existence of the eavesdropper, whereas if they use $S^\prime$ they can.

\section{Summary}
\label{SUM}

We present a new approach to simulate
quantum cryptography protocols using event-based processes.
The main feature of this approach is that it
simulates the transmission of the individual bits by an event-based process.
The algorithm that generates the events does not solve any quantum mechanical equation,
thereby circumventing the fundamental problems arising from the quantum measurement paradox.
Our simulation data for the BB84 and the Ekert protocol are, in all respects, in excellent agreement
with the theoretical expectations.
Extending the simulation method to account for effects such as depolarization by the
medium (fibers,air) and noise is left for future research.

\begin{acknowledgments}
We thank K. De Raedt, A. Keimpema, K. Michielsen, and S. Yuan for fruitful discussions.
\end{acknowledgments}

%
%

\appendix
\section{Wigner inequality}
\label{AppendixA}

We consider a two-particle system (particle $A$ and $B$)
and a pair of instruments that can measure a two-valued variable on each particle.
The two possible values of the observed variable are taken to be $\pm1$.
Each instrument has a range of settings.
For applications to quantum cryptography, it is sufficient
to consider the special case for which
particle $A$ is detected using one of the two settings $\phi_{A,1}$ and $\phi_{A,2}$,
and particle $B$ is detected using the two settings $\phi_{B,1}=\phi_{A,1}$ and $\phi_{B,2}$.
Each setting corresponds to a particular orientation of the apparatus that measures
the polarization.
It is assumed that the two observed results for a pair of particles are always opposite
if the two instruments have the same setting.
The Wigner inequality
\begin{equation}
P_{++}(\phi_{A,1},\phi_{B,2})+P_{++}(\phi_{A,2},\phi_{B,1})-P_{++}(\phi
_{A,2},\phi_{B,2})\geqslant0.
 \label{A0}
\end{equation}
is a convenient tool to characterize the correlation between the
results of the measurements on particles $A$ and $B$.

Proof: For any combination of settings, for example, $\phi_{A,1}$ and $\phi_{B,2}$,
the frequency of obtaining $+1$ on both sides is given by
\begin{equation}
F_{++}(\phi_{A,1},\phi_{B,2})=\frac{N_{++}(\phi_{A,1},\phi_{B,2})}{N}%
,
 \label{AF}
\end{equation}
where $N_{++}(\phi_{A,1},\phi_{B,2})$ denotes the number of events for
which both instruments yield $+1$ and $N$ is the total number of events.
For a different combination of the settings, the value of $N$ is assumed to be the same
(= ideal experiment assumption).
We now show that
\begin{eqnarray}
N_{++}(\phi_{A,1},\phi_{B,2})&+&N_{++}(\phi_{A,2},\phi_{B,1})
\nonumber \\
&-&N_{++}(\phi_{A,2},\phi_{B,2})\geqslant0,
 \label{A1}
\end{eqnarray}
holds under the conditions mentioned earlier.

Let us denote by $\phi_{A,1}^{(n)}$ and $\phi_{B,2}^{(n)}$ the results
recorded for the $i$th pair using the settings $\phi_{A,1}$ and $\phi_{B,2}$
Then $N_{++}(\phi_{A,1},\phi_{B,2})$ can be written as
\begin{equation}
N_{++}(\phi_{A,1},\phi_{B,2})=\sum_{n=1}^{N}\frac{1+\phi_{A,1}^{(n)}}{2}
\frac{1+\phi_{B,2}^{(n)}}{2}.  \label{countrelation}
\end{equation}

If the settings of the two instruments are such that
$\phi _{A,1}=\phi _{B,1}$ then we have $\phi _{A,1}^{(n)}=-\phi _{B,1}^{(n)}$
and
\begin{equation}
N_{++}(\phi _{A,1},\phi _{B,1})=\sum_{n=1}^{N}\frac{1+\phi _{A,1}^{(n)}}{2} \frac{1+\phi _{B,1}^{(n)}}{2}=0.
\end{equation}
Hence, instead of proving Eq.~(\ref{A1}), we can equally well prove that
\begin{eqnarray}
N_{++}(\phi _{A,1},\phi _{B,2}) &+&N_{++}(\phi _{A,2},\phi_{B,1})-N_{++}(\phi _{A,2},\phi _{B,2})
\notag \\
&-&N_{++}(\phi _{A,1},\phi _{B,1})\geqslant 0.
 \label{A2}
\end{eqnarray}

Substituting Eq.~(\ref{countrelation}) into Eq.~(\ref{A2})
we obtain
\begin{eqnarray}
\sum_{n=1}^{N}
&[
 ( 1+\phi_{A,1}^{(n)})(1+\phi_{B,2}^{(n)})
+( 1+\phi_{A,2}^{(n)})(1+\phi_{B,1}^{(n)})
\nonumber \\
&-( 1+\phi_{A,2}^{(n)}) (1+\phi_{B,2}^{(n)})
-( 1+\phi_{A,1}^{(n)}) (1+\phi_{B,1}^{(n)})
]
\nonumber \\
&
\geqslant0.
\end{eqnarray}
After simplification, we find
\begin{equation}
\sum_{n=1}^{N}
(\phi_{A,2}^{(n)}-\phi_{A,1}^{(n)}) (\phi_{B,2}^{(n)}-\phi_{B,1}^{(n)})
\geqslant0.
 \label{A3}
\end{equation}
Making use of the assumption that $\phi_{A,1}^{(n)}=-\phi_{B,1}^{(n)}$,
we find that
$( \phi_{A,2}^{(n)}-\phi_{A,1}^{(n)})t( \phi_{B,2}^{(n)}-\phi_{B,1}^{(n)}) \geqslant0$.
Hence, inequality Eq.~(\ref{A3}) holds and so do inequalities Eqs.~(\ref{A2}) and (\ref{A1}).

In the proof of Eq.~(\ref{A1}), it has been assumed that the observations
$\{\phi_{A,1}^{(n)}|n=1,\ldots,N\}$,
$\{\phi_{A,2}^{(n)}|n=1,\ldots,N\}$, $\{\phi_{B,1}^{(n)}|n=1,\ldots,N\}$,
and $\{\phi_{B,2}^{(n)}|n=1,\ldots,N\}$
do not depend on whether we record
$N_{++}(\phi _{A,1},\phi _{B,2})$, $N_{++}(\phi _{A,2},\phi_{B,1})$, or $N_{++}(\phi _{A,2},\phi _{B,2})$.
In a computer simulation, it is a simple matter to satisfy this assumption because we have perfect
control over the pseudo-random numbers that are used to generate the events but in
real experiments, the validity of this assumption cannot be taken for granted.

Having shown that
\begin{eqnarray}
F_{++}(\phi_{A,1},\phi_{B,2})&+&F_{++}(\phi_{A,2},\phi_{B,1})
\nonumber \\
&-&F_{++}(\phi_{A,2},\phi_{B,2})\geqslant0,
 \label{A4}
\end{eqnarray}
and assuming that the observations
$\{\phi_{A,1}^{(n)}|n=1,\ldots,N\}$,
$\{\phi_{A,2}^{(n)}|n=1,\ldots,N\}$, $\{\phi_{B,1}^{(n)}|n=1,\ldots,N\}$,
and $\{\phi_{B,2}^{(n)}|n=1,\ldots,N\}$
are independent random variables,
we may invoke the law of large numbers~\cite{GRIM95} to
argue that for $N\rightarrow\infty$,
$F_{++}(\phi_{A,1},\phi_{B,2})\rightarrow P_{++}(\phi_{A,1},\phi_{B,2})$
with probabilty one. Under these assumptions, the Wigner inequality Eq.~(\ref{A0}) holds.

\section{Modified Wigner inequality}
\label{AppendixB}

The proof of the modified Wigner inequality
\begin{eqnarray}
P_{++}(\phi _{A,1},\phi _{B,2}) &&+P_{++}(\phi _{A,2},\phi
_{B,1})+P_{--}(\phi _{A,1},\phi _{B,1})  \notag \\
&&-P_{++}(\phi _{A,2},\phi _{B,2})\geqslant 0,
\label{B0}
\end{eqnarray}
is very similar to the proof of the original Wigner inequality.
The essential difference is that the modified Wigner inequality
holds if we drop the assumption of perfect anticorrelation.

Adopting the same strategy as in Appendix~\ref{AppendixA},
we have to prove that
\begin{eqnarray}
N_{++}(\phi _{A,1},\phi _{B,2}) &+N_{++}(\phi _{A,2},\phi
_{B,1})+N_{--}(\phi _{A,1},\phi _{B,1})  \notag \\
&-N_{++}(\phi _{A,2},\phi _{B,2})\geqslant 0.
\label{B1}
\end{eqnarray}
Using Eq.~(\ref{countrelation}),
we can rewrite Eq.~(\ref{B1}) as
\begin{eqnarray}
\sum_{n=1}^{N}
&[
( 1+\phi_{A,1}^{(n)})( 1+\phi_{B,2}^{(n)})
+( 1+\phi_{A,2}^{(n)}) (1+\phi_{B,1}^{(n)})
\nonumber \\
&+( 1-\phi_{A,1}^{(n)}) (1-\phi_{B,1}^{(n)})
-( 1+\phi_{A,2}^{(n)}) (1+\phi_{B,2}^{(n)}) ]
\nonumber \\
&\geqslant0.
\label{B2}
\end{eqnarray}
After simplification, we find
\begin{equation}
\sum_{n=1}^{N}
[ 2+\phi_{B,2}^{(n)}( \phi_{A,1}^{(n)}-\phi_{A,2}^{(n)}) +\phi_{B,1}^{(n)}(\phi_{A,1}^{(n)}+\phi_{A,2}^{(n)})] \geqslant0.
\end{equation}
It is easy to see that
\begin{equation}
2+\phi _{B,2}^{(n)}( \phi _{A,1}^{(n)}-\phi _{A,2}^{(n)}) +\phi
_{B,1}^{(n)}( \phi _{A,1}^{(n)}+\phi _{A,2}^{(n)}) \geqslant 0,
\end{equation}
always holds.
Hence, inequality Eq.~(\ref{B1}) holds. Invoking the same
arguments that were used to replace frequencies by probabilities
in Appendix~\ref{AppendixA}, it then follows that
inequality Eq.~(\ref{B0}) holds.

\section{Probabilistic treatment}
\label{AppendixC}

If we replace the deterministic sequence of
pseudo-random numbers that we use in the computer simulations
by the mathematical concept of logically independent random variables,
as defined in the (Kolmogorov) theory of probabilitity~\cite{GRIM95,JAYN03},
we can readily obtain analytical expressions for the expectation values
that we compute with the simulation model.
This then allows us to analyze the event-based simulation of Ekert's protocol
analytically. In particular, we will prove that
for both the deterministic model of the polarized beam splitter (DP) and
the probabilistic model of the polarized beam splitter (PP),
the event-by-event simulation reproduces \textit{exactly}
the two-particle probability Eq.~(\ref{countp++}) of quantum theory.

We start by assuming that there exists a probability,
denoted by $P(x_{1},x_{2},t_{1},t_{2}|\alpha ,\beta )$, to observe the data
$\left\{x_{1},t_{1}\right\} $ and $\left\{ x_{2},t_{2}\right\} $ for fixed
orientations $\left\{ \alpha ,\beta \right\} $ on both observation stations.

\begin{widetext}

As a first step, let us express the probability for observing
the data $\left\{ x_{1},x_{2},t_{1},t_{2}\right\} $ as an integral over the
mutually exclusive events $\xi _{1},\xi _{2}$. According to the rules of
probability theory \cite{GRIM95,JAYN03}, we have
\begin{equation}  \label{pepr0}
P(x_1,x_2,t_1,t_2|\alpha,\beta)= \frac{1}{4\pi^2}\int_0^{2\pi} \int_0^{2\pi}
P(x_1,x_2,t_1,t_2|\alpha,\beta,\xi_1,\xi_2) P(\xi_1,\xi_2|\alpha,\beta)
d\xi_1 d\xi_2,
\end{equation}
where $\xi _{1},\xi _{2}$ denotes the two-dimensional unit vector
representing the polarization. Starting from the exact representation
Eq.~(\ref{pepr0}), we may now use the knowledge that in our simulation model
(but not necessarily in experiment), for each event, the values of $%
\{x_1,x_2,t_1,t_2\}$ are logically independent of each other and that the
values of $\{x_1,t_1\}$ ($\{x_2,t_2\}$) are also logically independent of $%
\beta$ and $\xi_2$ ($\alpha$ and $\xi_1$)). Thus, we may write
\begin{eqnarray}  \label{pepr1}
P(x_1,x_2,t_1,t_2|\alpha,\beta) &=& \frac{1}{4\pi^2}\int_0^{2\pi}
\int_0^{2\pi} P(x_1,t_1|x_2,t_2,\alpha,\beta,\xi_1,\xi_2)
P(x_2,t_2|\alpha,\beta,\xi_1,\xi_2) P(\xi_1,\xi_2|\alpha,\beta) d\xi_1 d\xi_2
\notag \\
&=& \frac{1}{4\pi^2}\int_0^{2\pi} \int_0^{2\pi} P(x_1,t_1|\alpha,\xi_1)
P(x_2,t_2|\beta,\xi_2) P(\xi_1,\xi_2|\alpha,\beta) d\xi_1 d\xi_2   \notag \\
&=& \frac{1}{4\pi^2}\int_0^{2\pi} \int_0^{2\pi} P(x_1|\alpha,\xi_1)
P(t_1|\alpha,\xi_1) P(x_2|\beta,\xi_2) P(t_2|\beta,\xi_2)
P(\xi_1,\xi_2|\alpha,\beta) d\xi_1 d\xi_2   \notag \\
&=& \frac{1}{4\pi^2}\int_0^{2\pi} \int_0^{2\pi} P(x_1|\alpha,\xi_1)
P(t_1|\alpha,\xi_1) P(x_2|\beta,\xi_2) P(t_2|\beta,\xi_2) P(\xi_1,\xi_2)
d\xi_1 d\xi_2 ,
\end{eqnarray}
where, in the last step, we used the knowledge that in our simulation model,
the values of $\xi_1$ and $\xi_2$ are logically independent of $\alpha$ or $%
\beta$. Note that Eq.~(\ref{pepr1}) gives the exact probabilistic
description of our simulation model.

The mathematical structure of Eq.~(\ref{pepr1}) is the same as the one that
is used in the derivation of Bell's results.
Thus, if we would continue along the same line as in Bell's work
the model defined by Eq.~(\ref{pepr1}) cannot produce the correlation of the singlet state.
However, the real factual situation in the experiment is different:
The events are selected using a time window $W$ that the experimenters try
to make as small as possible~\cite{WEIH00}. Accounting for the time window,
that is multiplying Eq.~(\ref{pepr1}) by the step function (see Eq.~(\ref{Cxy})) and integrating
over all $t_1$ and $t_2$, the expression for the probability for observing
the event $(x_1,x_2)$ reads
\begin{eqnarray}
P(x_{1},x_{2}|\alpha ,\beta ) &=&\frac{\int_{0}^{2\pi }\int_{0}^{2\pi
}P(x_{1}|\alpha ,\xi _{1})P(x_{2}|\beta ,\xi _{2})w(\alpha ,\beta ,\xi
_{1},\xi _{2},W)P(\xi _{1},\xi _{2})d\xi _{1}d\xi _{2}}{\sum_{x_{1},x_{2}=%
\pm 1}\int_{0}^{2\pi }\int_{0}^{2\pi }P(x_{1}|\alpha ,\xi _{1})P(x_{2}|\beta
,\xi _{2})w(\alpha ,\beta ,\xi _{1},\xi _{2},W)P(\xi _{1},\xi _{2})d\xi
_{1}d\xi _{2}}  \notag  \label{pepr2} \\
&=&\frac{\int_{0}^{2\pi }\int_{0}^{2\pi }P(x_{1}|\alpha ,\xi
_{1})P(x_{2}|\beta ,\xi _{2})w(\alpha ,\beta ,\xi _{1},\xi _{2},W)P(\xi
_{1},\xi _{2})d\xi _{1}d\xi _{2}}{\int_{0}^{2\pi }\int_{0}^{2\pi }w(\alpha
,\beta ,\xi _{1},\xi _{2},W)P(\xi _{1},\xi _{2})d\xi _{1}d\xi _{2}},
\end{eqnarray}%
where the weight function in our simulation is
\begin{equation}
w(\alpha ,\beta ,\xi _{1},\xi _{2},W)=\frac{1}{T_{1}T_{2}}%
\int_{0}^{T_{1}}dt_{1}\int_{0}^{T_{2}}dt_{2}\,\Theta (W-|t_{1}-t_{2}|).
\label{weightfunction}
\end{equation}%
with $T_{1}=T_{0}|\sin 2(\alpha -\xi _{1})|^{d}$ and $T_{2}=T_{0}|\sin
2(\beta -\xi _{2})|^{d}$, and the time delays $t_{i}$ are distributed
uniformly over the interval $[0,T_{i}]$. The weight function will be less
than one unless $W$ is larger than the range of $(t_{1},t_{2})$. The
integrals in Eq.~(\ref{weightfunction}) can be worked out analytically,
yielding
\begin{eqnarray}
w(\alpha ,\beta ,\xi _{1},\xi _{2},W)=\frac{1}{4T_{1}T_{2}}
&[&T_{1}^{2}+T_{2}^{2}+2(T_{1}+T_{2})W+(W-T_{1})|W-T_{1}|+(W-T_{2})|W-T_{2}|
\notag \\
&&-(W-T_{1}+T_{2})|W-T_{1}+T_{2}|-(W+T_{1}-T_{2})|W+T_{1}-T_{2}|\;\;].
\label{sweightfunction}
\end{eqnarray}

We now consider the specific case of the PP and the DP model, respectively.
The PP reproduces Malus law for a single polarizer, that is
the probability distributions $P(x_{1}|\alpha ,\xi _{1})$ and $P(x_{2}|\beta
,\xi _{2})$ can be written as
\begin{eqnarray}  \label{pepr6}
P(x_{1}|\alpha ,\xi _{1}) &=&\frac{1+x_{1}\cos 2(\alpha -\xi _{1})}{2}, 
\notag \\
P(x_{2}|\beta ,\xi _{2}) &=&\frac{1+x_{2}\cos 2(\beta -\xi _{2})}{2}.
\label{ppmodel}
\end{eqnarray}
Let us now consider the case of Ekert's protocol and specialize to the case
that the source emits particles with opposite polarization
$P(\xi _{1},\xi_{2})=\delta (\xi _{1}+\pi /2-\xi _{2})P(\xi _{1})$
with $P(\xi _{1})$ being
a uniform distribution. If $d=0$ and $W\leq T_{0}$, we have $w(\alpha ,\beta
,\xi _{1},\xi _{2},W)=(2T_{0}-W)W/T_{0}^{2}$. Likewise, if $W>T_{0}$, $%
w(\alpha ,\beta ,\xi _{1},\xi _{2},W)=1$. Therefore, if $W>T_{0}$ or $d=0$,
we have
\begin{eqnarray}
P(x_{1},x_{2}|\alpha ,\beta ) &=&\frac{\int_{0}^{2\pi }\int_{0}^{2\pi
}P(x_{1}|\alpha ,\xi _{1})P(x_{2}|\beta ,\xi _{2})P(\xi _{1},\xi _{2})d\xi
_{1}d\xi _{2}}{\int_{0}^{2\pi }\int_{0}^{2\pi }P(\xi _{1},\xi _{2})d\xi
_{1}d\xi _{2}}  \notag  \label{pepr11} \\
&=&\frac{1}{8\pi }\int_{0}^{2\pi }(1+x_{1}\cos 2(\alpha -\xi ))(1-x_{2}\cos
2(\beta -\xi ))d\xi  \notag \\
&=&\frac{2-x_{1}x_{2}\cos 2(\alpha -\beta )}{8},
\label{pd0pmodel}
\end{eqnarray}
and, more specifically,
\begin{equation}
P_{++}(\alpha ,\beta )=\frac{2-\cos 2(\alpha -\beta )}{8}.
\label{pd0++pmodel}
\end{equation}
The corresponding expression of the Wigner parameter $S$ reads
\begin{equation}
\text{\ }S=\frac{2-\cos 2\theta +\cos 4\theta }{8}.  \label{sd0pmodel}
\end{equation}
From Eqs.~(\ref{pd0pmodel}), (\ref{pd0++pmodel}) and (\ref{sd0pmodel}),
it is clear that if we ignore the time-tag information, the two-particle
probability takes the form of the hidden variable models considered by Bell~\cite{BELL93},
and the event-based model cannot reproduce the results of quantum theory.

Next, we take into account the time-tag information and, to be able to obtain a closed-form expression,
we focus on the limit $W\rightarrow 0$.
Then, Eq.~(\ref{sweightfunction}) reduces to
\begin{equation}
w(\alpha ,\beta ,\xi _{1},\xi _{2},W\rightarrow 0)=\frac{2W}{\max
(T_{1},T_{2})}+\mathcal{O}(W^{2}).  \label{pepr7}
\end{equation}

Inserting Eq.~(\ref{pepr7}) and $P(\xi _{1},\xi_{2})=\delta (\xi _{1}+\pi /2-\xi _{2})$
into Eq.~(\ref{pepr2}) we find
\begin{eqnarray}
P(x_{1},x_{2}|\alpha ,\beta ) &=&\frac{\int_{0}^{2\pi }\int_{0}^{2\pi
}P(x_{1}|\alpha ,\xi _{1})P(x_{2}|\beta ,\xi _{2})P(\xi _{1},\xi _{2})\max
(|\sin 2(\xi _{1}-\alpha )|,|\sin 2(\xi _{2}-\beta )|)^{-d}d\xi _{1}d\xi _{2}%
}{\int_{0}^{2\pi }\int_{0}^{2\pi }P(\xi _{1},\xi _{2})\max (|\sin 2(\xi
_{1}-\alpha )|,|\sin 2(\xi _{2}-\beta )|)^{-d}d\xi _{1}d\xi _{2}}  \notag \\
&=&\frac{1}{4} \frac{\int_{0}^{2\pi }(1+x_{1}\cos 2(\alpha -\xi
))(1-x_{2}\cos 2(\beta -\xi ))\max (|\sin 2(\xi -\alpha )|,|\sin 2(\xi
-\beta )|)^{-d}d\xi }{\int_{0}^{2\pi }\max (|\sin 2(\xi -\alpha )|,|\sin
2(\xi -\beta )|)^{-d}d\xi }.  \label{pd4pmodel}
\end{eqnarray}

For $d=4$ the integrals in Eq.~(\ref{pd4pmodel}) can be worked out analytically
and the result of this excercise reads
\begin{equation}
P(x_{1},x_{2}|\alpha ,\beta )=\frac{1-x_{1}x_{2}\cos 2(\alpha -\beta )%
}{4},  \label{pd41pmodel}
\end{equation}
yielding
\begin{equation}
P_{++}(\alpha ,\beta )=\frac{1}{2}\sin ^{2}(\alpha -\beta ),
\label{pd4++pmodel}
\end{equation}%
in exact agreement with the expression for a system of two $S=1/2$
quantum objects in the singlet state.

The analytical results for the deterministic model of
the polarizing beam splitter can be derived in the same manner:
We only have to change the specific expression for the probability distribution
of a single polarizer.
In the DP model, the values of $x_{1}$ and $x_{2}$ are determined by the $sign$ function.
The corresponding probability distributions can be written as
\begin{eqnarray}
P(x_{1}|\alpha ,\xi _{1}) &=&\Theta \left( x_{1}\cos 2(\alpha -\xi
_{1})\right) ,  \notag \\
P(x_{2}|\beta ,\xi _{2}) &=&\Theta \left( x_{2}\cos 2(\beta -\xi
_{2})\right) .  \label{pdmodel}
\end{eqnarray}
Now we study the same specific cases as we did earlier.
First we consider the case for which $W>T_{0}$ or $d=0$.
Then, we have
\begin{eqnarray}
P(x_{1},x_{2}|\alpha ,\beta ) &=&\frac{\int_{0}^{2\pi }\int_{0}^{2\pi
}P(x_{1}|\alpha ,\xi _{1})P(x_{2}|\beta ,\xi _{2})P(\xi _{1},\xi _{2})d\xi
_{1}d\xi _{2}}{\int_{0}^{2\pi }\int_{0}^{2\pi }P(\xi _{1},\xi _{2})d\xi
_{1}d\xi _{2}}  \notag \\
&=&\frac{1}{2\pi }\int_{0}^{2\pi }\Theta \left( x_{1}\cos 2(\alpha -\xi
)\right) \Theta \left( -x_{2}\cos 2(\beta -\xi )\right) d\xi ,
\label{pd0dmodel}
\end{eqnarray}%
yielding
\begin{equation}
P_{++}(\alpha ,\beta )=\frac{1}{2}-\left\vert \frac{\left\vert \alpha -\beta
\right\vert }{\pi }-\frac{1}{2}\right\vert .  \label{pd0++dmodel}
\end{equation}
Second, we consider the limit $W\rightarrow 0$ and find that for fixed $(\alpha ,\beta )$,
the probability for observing the event $(x_{1},x_{2})$ is given by
\begin{eqnarray}
P(x_{1},x_{2}|\alpha ,\beta ) &=&\frac{\int_{0}^{2\pi }\int_{0}^{2\pi
}P(x_{1}|\alpha ,\xi _{1})P(x_{2}|\beta ,\xi _{2})\max (|\sin 2(\xi
_{1}-\alpha )|,|\sin 2(\xi _{2}-\beta )|)^{-d}P(\xi _{1},\xi _{2})d\xi
_{1}d\xi _{2}}{\int_{0}^{2\pi }\int_{0}^{2\pi }\max (|\sin 2(\xi _{1}-\alpha
)|,|\sin 2(\xi _{2}-\beta )|)^{-d}P(\xi _{1},\xi _{2})d\xi _{1}d\xi _{2}} 
\notag \\
&=&\frac{\int_{0}^{2\pi }\Theta \left( x_{1}\cos 2(\alpha -\xi )\right)
\Theta \left( -x_{2}\cos 2(\beta -\xi )\right) \max (|\sin 2(\xi -\alpha
)|,|\sin 2(\xi -\beta )|)^{-d}d\xi }{\int_{0}^{2\pi }\max (|\sin 2(\xi
-\alpha )|,|\sin 2(\xi -\beta )|)^{-d}d\xi }.
\label{pd2dmodel}
\end{eqnarray}
Writing $\theta =\beta -\alpha$ and putting $d=2$, we find
that the probability $P_{++}(\alpha,\beta )$ is given by
\begin{eqnarray}
P_{++}(\alpha ,\beta ) &=&\frac{1}{2}\frac{\int_{3\pi /4}^{3\pi /4+\theta /2}|\sin 2\xi
|^{-d}d\xi +\int_{3\pi /4+\theta /2}^{3\pi /4+\theta }|\sin 2(\xi -\theta
)|^{-d}d\xi }{\int_{\theta /2}^{\pi /4+\theta /2}|\sin 2\xi |^{-d}d\xi
+\int_{\pi /4+\theta /2}^{\pi /2+\theta /2}|\sin 2(\xi -\theta )|^{-d}d\xi }
\notag \\
&=&\frac{1}{2}\sin ^{2}\theta  \label{pd2++dmodel},
\end{eqnarray}%
in exact agreement with the expression for a system of two $S=1/2$
quantum objects in the singlet state.

\end{widetext}

\bibliography{qkd}

\end{document}